\documentclass[10pt,conference,a4paper,oneside,twocolumn]{IEEEtran}

\usepackage{amsthm}
\usepackage{amsmath,amssymb}
\usepackage{psfrag}
\usepackage{pifont}
\usepackage{graphicx}
\usepackage[nocompress]{cite}
\usepackage{fancyhdr}
\graphicspath{{drawings/}{pics/}}
\usepackage{color}
\usepackage[normalem]{ulem}
\usepackage{algorithmic}
\usepackage[ruled]{algorithm}

\definecolor{mygray}{rgb}{0.6,0.6,0.6}

\def\ve#1{{\mathchoice{\mbox{\boldmath$\displaystyle #1$}}%
              {\mbox{\boldmath$\textstyle #1$}}%
              {\mbox{\boldmath$\scriptstyle #1$}}%
              {\mbox{\boldmath$\scriptscriptstyle #1$}}}}
\DeclareSymbolFont{AMSb}{U}{msb}{m}{n}
\DeclareSymbolFontAlphabet{\mathbb}{AMSb}

\def\sign{\mathrm{sign}}

\def\e{\mathrm{ e}}

\def\Pr{\mathrm{Pr}}
\def\f{\mathsf{f}}

\def\LLR{\mathrm{LLR}}
\def\LLRmax{\LLR_{\mathsf{max}}}
\def\Lmax{\Lambda_{\mathsf{max}}}
\def\GLRT{\mathsf{MSDD}}
\newcommand {\dx} {~\mathrm{d}}

\newcommand {\pTX} {p^{\mathsf{TX}}}     
\newcommand {\T} {T}     
\newcommand {\Ti} {T_{\mathsf{i}}}     
\newcommand {\hCH} {h^{\mathsf{CH}}}		
\newcommand {\hRX} {h^{\mathsf{RX}}}		

\newcommand {\Rstop} {R_{\mathsf{stop}}} 

\newcommand{\pathMw}[1]{\Delta_{#1}}

\def\dB{\mathrm{dB}}
\def\ns{\,\mathrm{ns}}
\def\GHz{\,\mathrm{GHz}}

\newcommand {\Eb} {E_{\mathrm{b}}}     
\def\No{N_0}
\def\EbNo{{\Eb}/{\No}}

\def\BER{\mathrm{BER}}

\newcommand{\argmin}{\mathop{\mathrm{argmin}}}

\renewcommand{\algorithmicif}{\textbf{{if}}}

\renewcommand{\algorithmicwhile}{\textbf{{{while}}}}

\definecolor{commentgray}{rgb}{.3,.3,.3}

\newcommand{\CODE}[1]{\textbf{#1}}

\definecolor{highlightgray}{rgb}{.9,.9,.9}
\newcommand{\HIGHLIGHT}[1]{\colorbox{highlightgray}{#1}}
\newcommand{\MANINDENT}{\hspace*{1em}}
\newcommand{\FUNCTION}[3]{\CODE{function} [#3] = \texttt{#1}(#2)}
\newcommand{\FUNCTIONCALL}[3]{[#3] := \texttt{#1}(#2)}

\newcommand{\IFONELINE}[2]{\STATE \algorithmicif ~ #1 {\large\{} #2 {\large\}}}
\newcommand{\WHILEONELINE}[2]{\algorithmicwhile ~ #1 {\large\{} #2 {\large\}}}

\newcommand{\veat}{\ve{\tilde{a}}}
\DeclareFontFamily{OT1}{phv}{}
\DeclareFontShape{OT1}{phv}{m}{n}{ <-> [0.95] phvr8t }{}

\title{\huge Soft-Output Sphere Decoder for Multiple-Symbol Differential Detection of Impulse-Radio Ultra-Wideband}

\IEEEoverridecommandlockouts
\author{
\IEEEauthorblockN{Andreas~Schenk and Robert~F.H.~Fischer}
\IEEEauthorblockA{Lehrstuhl f{\"u}r Informations{\"u}bertragung, Universit{\"a}t Erlangen--N{\"u}rnberg, Erlangen, Germany}
Email: \texttt{\{schenk,fischer\}@lnt.de}

\thanks{This work was supported by the Deutsche Forschungsgemeinschaft (DFG) within the framework UKoLoS under grant FI 982/3-1.}
}
\begin{document}
\pagestyle{fancy} 
\fancyhead[RE,LO]{\color{mygray}Preprint of ISIT paper \#1061 --- Andreas Schenk, Robert~F.H.~Fischer:\\Soft-Output Sphere Decoder for Multiple-Symbol Differential Detection of Impulse-Radio Ultra-Wideband}
\fancyfoot[LE,RO]{\color{mygray}Andreas~Schenk and Robert~F.~H.~Fischer, \today}

\maketitle

\begin{abstract}
Power efficiency of noncoherent receivers for impulse-radio ultra-wideband (IR-UWB) transmission systems can significantly be improved, on the one hand, by employing multiple-symbol differential detection (MSDD), and, on the other hand, by providing reliability information to the subsequent channel decoder.
In this paper, we combine these two techniques.
Incorporating the computation of the soft information into a single-tree-search sphere decoder (SD), the application of this soft-output MSDD in a typical IR-UWB system imposes only a moderate complexity increase at, however, improved performance over hard-output MSDD, and in particular, over conventional symbol-by-symbol noncoherent differential detection.
\end{abstract}

%
\section{Introduction}
\label{sec:intro}
%
Impulse-radio ultra-wideband (IR-UWB) is widely considered as a promising technique for low-power low-cost short-range wireless communication systems.
One of the main reasons for this is its potential to employ noncoherent, hence low-complexity, receivers even in dense multipath propagation scenarios, where channel estimation required for coherent detection would be overly complex due to the high multipath resolution and relatively large delay spread of UWB signals.

The performance penalty between coherent and noncoherent detection in power efficiency, i.e., in the required signal-to-noise ratio to guarantee a certain bit error rate (BER), can be closed by replacing conventional symbol-by-symbol noncoherent detection with a joint detection of a block of symbols, i.e., performing multiple-symbol differential detection (MSDD) \cite{me:IZS2010,UWB:Guo:MSDD}.
In \cite{UWB:Lottici:MSDD} it has been shown that the underlying tree search problem is efficiently solved using the sphere decoder (SD) algorithm.
However, noncoherent (MSDD-based) IR-UWB receiver design has mainly considered uncoded transmission systems, cf., e.g., \cite{UWB:Chao:TR,UWB:Lottici:MSDD,UWB:Guo:MSDD}.

In this paper, we consider coded IR-UWB transmission employing differentially encoded BPSK (also known as differential transmitted reference (DTR)).
To keep transmitter and receiver design simple, we restrict to the conventional serial concatenation of modulation and coding at transmitter, and detection and decoding at receiver side, i.e., restrain to the bit-interleaved coded modulation (BICM) philosophy.
Employing coding, power efficiency can significantly be improved by delivering reliability information to the soft-input channel decoder \cite{book:WozencraftJacobs:Principles}.
Borrowing from techniques recently introduced for SD-based soft output generation in multiple-input/multiple-output (MIMO) systems \cite{SISOSD:StuderBurgBolcskei:SOSD}, we extend the SD-based MSDD algorithm presented in \cite{UWB:Lottici:MSDD} to incorporate also soft output computation.

To this end, in Section~\ref{sec:ir}, we derive log-likelihood ratios (LLR) for MSDD of IR-UWB based on generalized-likelihood ratio testing (GLRT), and formulate their computation as a tree search problem.
Using the soft-output sphere decoder (SOSD) the  LLRs can be found in a single tree search, thus receiver complexity is increased only moderately, especially in comparison to repeated-tree-search approaches, cf. \cite{SISOSD:StuderBurgBolcskei:SOSD}.

In Section~\ref{sec:results}, we investigate the performance of the proposed MSDD-based soft-output IR-UWB receiver and study the tradeoff between performance and complexity obtained by adjusting the channel code, the MSDD block size, and techniques for SD complexity reduction, such as a stopping criterion \cite{me:ICUWB09} and LLR clipping \cite{SISOSD:StuderBurgBolcskei:SOSD}.
We conclude with a summary in Section~\ref{sec:conclusions}.

%
\section{Impulse-Radio Ultra-Wideband Transmission}
\label{sec:ir}
\subsection{System Model}
\label{sec:ir:system}
\begin{figure*}[!t]
\centering
{\footnotesize
        \input{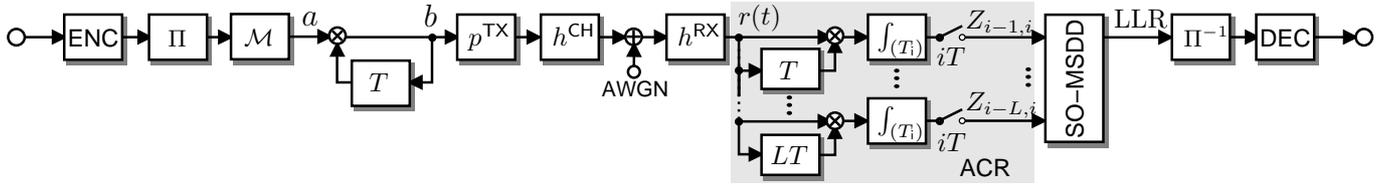}}
\caption{Block diagram of the coded IR-UWB transmission system with $L$-branch ACR and soft-output MSDD.}
\vspace*{-3mm}
\label{fig:UWB:ACR}
\end{figure*}

The receive signal of differentially encoded BPSK IR-UWB is given as (cf.\ Fig.~\ref{fig:UWB:ACR})
\begin{align}
        r(t) = \sum_{i=0}^{+\infty} b_i p (t-i\T) + n(t)
        \label{eq:UWB:RX}
\end{align}
where $b_i$ are the differentially encoded, interleaved ($\Pi$) and mapped ($\mathcal{M}$) output symbols $a_i\in\mathcal{A} = \{\pm1\}$ of a channel encoder, such that $b_i = b_0 \prod_{k=1}^{i}a_k$, with $b_0 = 1$, and $\T$ is the symbol duration.
The receive pulse shape $p(t)$ is obtained from the convolution of transmit pulse, receive filter, and channel impulse response, i.e., $p(t) = \hCH(t) * \hRX(t) * \pTX(t)$.
The pulse energy is normalized to one and thus, the energy per bit is given by $\Eb = 1$.
$n(t)$ is white Gaussian noise of two-sided power-spectral density $\No/2$ filtered by $\hRX(t)$.
To preclude inter-symbol interference, the symbol duration $\T$ is chosen sufficiently large, such that each pulse has decayed before the next pulse is received.
Note that the usually applied frame structure used for time-hopping and code-division multiple access \cite{UWB:Win:IR,UWB:Win:TH} is not explicitly taken into account, as it can be regarded as additional linear block coding, or removed prior to further receive signal processing \cite{UWB:Lottici:MSDD}.

%
\subsection{Soft Output Generation}
The reliability information for a single information symbol, which is passed as soft input to the channel decoder, is calculated based on the observation of the receive signal in an $(L+1)$-symbols interval $0\leq t< (L+1)\T$
(without loss of generality we consider the interval starting at $t=0$).
The channel is assumed to be constant in this interval, which, in typical indoor UWB communication scenarios, is fulfilled especially for moderate $L$ \cite{UWB:Molisch:Channel}.

Due to the unknown statistics of the receive pulse shape $p(t)$, we revert to the GLRT approach, thus, in contrast to the ML criterion, include an explicit optimization over all finite-energy pulse shapes $\tilde{p}(t)$ of some assumed duration $\Ti$ \cite{UWB:Lottici:MSDD,UWB:Chao:TR,me:IZS2010}.
In terms of LLRs for the $i^{\mathrm{th}}$ symbol, $i=1,...,L$, this reads
\begin{align}
        \LLR_i &= \log\left( \frac{\max_{\tilde{p}(t)} \Pr\{\tilde{a}_i = +1 | r(t),\tilde{p}(t)\} }{\max_{\tilde{p}(t)} \Pr\{\tilde{a}_i = -1 | r(t),\tilde{p}(t)\}}\right)\;.
\label{eq:LLR_first}
\end{align}
Defining the vector of the information symbol hypotheses ${\veat} = [\tilde{a}_1, ...,\tilde{a}_L]$, applying Bayes' rule, and assuming independent i.i.d.\ information symbols, i.e., a-priori information is not considered as feedback from the channel decoder, we have 
\begin{align}
        \LLR_i &= \log\left( \frac{\mathop{\sum_{\veat\in\mathcal{A}^L}}_{,\tilde{a}_i = +1} \max_{\tilde{p}(t)} \f(r(t)|\veat,\tilde{p}(t))}
                                  {\mathop{\sum_{\veat\in\mathcal{A}^L}}_{,\tilde{a}_i = -1} \max_{\tilde{p}(t)} \f(r(t)|\veat,\tilde{p}(t))}\right)
\label{eq:LLR_second}
\end{align}
where $\f(r(t)|\veat,\tilde{p}(t))$ is the probability density function of the additive noise in (\ref{eq:UWB:RX}).
As $\f(r(t)|\veat,\tilde{p}(t)) \geq 0$, maximization and summation could be interchanged.

In a first step, we perform the maximization over the unknown pulse shape, applying similar steps as shown in \cite{UWB:Lottici:MSDD}.
This is summarized, very briefly, as follows.

We define the noise-free receive signal hypothesis $s(t|\veat,\tilde{p}(t)) = \sum_{i=0}^{L} b_0 \prod_{k=1}^{i} \tilde{a}_k \tilde{p}(t-i\T)$, including the  receive pulse shape hypothesis $\tilde{p}(t)$.
Since additive Gaussian noise is assumed, which is bandlimited by a receive filter with equivalent noise bandwidth $B_{\mathsf{eq}}$, and relying on the equivalence of distance in signal space and energy of the difference signal \cite{book:Proakis:DigitalCommunications}, we arrive at the joint probability density function
\begin{align}
        \f(r(t)|\veat,\tilde{p}(t))
      &= c_{\mathsf{f}} \cdot\e^{-\frac{1}{2\sigma_n^2}\int_0^{(L+1)\T} \left( r(t) - s(t|\veat,\tilde{p}(t))\right)^2\dx t}
\label{eq:pdf}
\end{align}
where $\sigma_n^2 = \No B_{\mathsf{eq}}$, and $c_{\mathsf{f}}$ is an irrelevant constant, which cancels in (\ref{eq:LLR_second}).
Applying variational calculus and using the fact that $\tilde{a}_i \in\{\pm1\}$, the optimizing pulse shape is found to be $p_{\mathsf{opt}}(t) = \frac{1}{L+1}\,\sum_{i=0}^{L} b_0\prod_{k=1}^{i}\tilde{a}_i\,r(t+i\T)$ (cf. \cite{UWB:Lottici:MSDD} for a similar derivation). Thus,
\begin{align}
      \max_{\tilde{p}(t)}\f(r(t)|\veat,\tilde{p}(t)) &= c_{\mathsf{f}} \cdot \e^{-\frac{-2\,\Gamma(\veat) - \sum_{i=0}^{L} Z_{i,i}}{2\sigma_n^2(L+1)} }
\label{eq:maxpdf}
\end{align}
where we defined
\begin{align}
        \Gamma(\veat) =  \sum_{i=1}^{L}\sum_{l=0}^{i-1}\prod_{k=l+1}^{i} \tilde{a}_k Z_{l,i}
\label{eq:Gamma}
\end{align}
and, for $i=1,...,L$, $l = 0,...,i-1$,
\begin{align}
        Z_{l,i} &= \int_{0}^{\Ti} r(t+i\T)\cdot r(t+l\T) \dx t\;.
\label{eq:definition_Z}
\end{align}

Using the result in (\ref{eq:maxpdf}), as well as applying the max-log approximation, (\ref{eq:LLR_second}) can be approximated by
\begin{align}
\LLR_i ={}&\frac{1}{\sigma_n^2(L+1)} \left[\mathop{\max_{\veat\in\mathcal{A}^L}}_{\tilde{a}_i = +1} \Gamma(\veat) - \mathop{\max_{\veat\in\mathcal{A}^L}}_{\tilde{a}_i = -1}\Gamma(\veat)\right]\;.
\label{eq:LLR_third}
\end{align}
%

%
\subsection{Autocorrelation-Based Detection}
\label{sec:ir:sodd}
Before we turn to an efficient implementation of the computation required for (\ref{eq:LLR_third}) based on the SD algorithm, we note that $Z_{l,i}$, defined in (\ref{eq:definition_Z}), corresponds to the output of an $L$-branch autocorrelation receiver (ACR, shown in Fig.~\ref{fig:UWB:ACR}) with delays being multiples of $\T$ and integration interval $\Ti$, set in the order of the expected receive pulse duration.
$Z_{l,i}$ represents the phase transition from $b_l$ to $b_i$ superposed by an ``information $\times$ noise'' and ``noise $\times$ noise'' term.
%
%

Clearly, restricting to a single-branch ACR, i.e., $L=1$, corresponds to symbol-by-symbol differential detection (DD).
In this case, it can directly be seen that the hard-quantized ACR output gives the DD estimate, i.e.,  $a_i^{\mathsf{DD}} = \sign\left(Z_{i-1,i}\right)$.
However, using (\ref{eq:LLR_third}) with $L=1$, also the rather intuitive result follows, to use the unquantized ACR output for soft-output DD, in particular $\LLR_i^{\mathsf{DD}} = \frac{1}{\sigma_n^2\cdot2} [2 Z_{i-1,i}]$.
%
%

%
\subsection{Soft-Output Sphere Decoder (SOSD)}
\label{sec:ir:sosd}
For efficient implementation based on the SD algorithm, we reformulate the maximization problems in (\ref{eq:LLR_third}) into equivalent minimization problems.
Since $\tilde{a}_i \in \{\pm1\}$, $\Gamma(\veat) \leq \sum_{i=1}^{L}\sum_{l=0}^{i-1}|Z_{l,i}|$ holds $\forall \veat$. Subtracting this upper bound from both objective functions in (\ref{eq:LLR_third}) yields
\begin{align}
        \LLR_i ={} \frac{1}{\sigma_n^2(L+1)} \left[
\mathop{\min_{\veat\in\mathcal{A}^L}}_{\tilde{a}_i = -1}  \Lambda(\veat) -\mathop{\min_{\veat\in\mathcal{A}^L}}_{\tilde{a}_i = +1}  \Lambda(\veat) \right]\;.
\label{eq:LLR_four}
\end{align}
%
Note that $\max(x-y) = -\min(y-x)$, and
\begin{align}
        \Lambda(\veat) = \sum_{i=1}^{L}\sum_{l=0}^{i-1} |Z_{l,i}|\left(1-\sign(Z_{l,i}) \prod_{k=l+1}^{i} \tilde{a}_k \right)\;.
\label{eq:Lambda}
\end{align}
Eventually, it can be seen that the addends of the outer sum in (\ref{eq:Lambda}) are always non-negative and depend solely on the first $i$ (preliminary) decisions of information symbols $\tilde{a}_k$, $k=1,...,i$.
This allows to check the decision metric componentwise, and thus each of the two minimization problems in (\ref{eq:LLR_four}) can be solved using the SD operating on an $L$-dimensional binary tree (see Fig.~\ref{fig:UWB:pseudocode} and also \cite{UWB:Lottici:MSDD} for details).

Clearly, one of the two minima in (\ref{eq:LLR_four}) is the GLRT-optimal metric
\begin{align}
        \Lambda^\GLRT = \Lambda(\ve{a}^\GLRT)= \min_{\veat\in\mathcal{A}^L} \Lambda(\veat)
\label{eq:}
\end{align}
whereas the other one is obtained from the corresponding counter hypothesis, i.e., the minimum metric with the restriction $\tilde{a}_i = -a_i^\GLRT$, such that
\begin{align}
        \Lambda_i^{\overline{\GLRT}} = 
\min_{\veat\in\mathcal{A}^L,\, \tilde{a}_i = -{a}_i^\GLRT} \Lambda(\veat)\;.
\label{eq:UWB:Lambdacounter}
\end{align}
Consequently, we have
\begin{align}
        \LLR_i = 
\frac{1}{\sigma_n^2(L+1)}\, \left[ a_i^\GLRT\,\left( \Lambda_i^{\overline{\GLRT}} - \Lambda^{{\GLRT}}\right) \right]
\;.
\label{eq:UWB:LLR_cases}
\end{align}
\subsubsection{Single-Tree-Search SOSD}
Calculating the LLRs resorts to finding the minimum of an unrestricted tree search, the corresponding sequence, and $L$ ``next-best'' minima.
One could solve these minimization problems subsequently by rerunning the SD for each counterhypothesis with correspondingly restricted search space.
This requires to run the SD $L+1$ times per block of $L$ information symbols, and hence, imposes a high complexity burden.

This can be alleviated by a modified SD algorithm, as introduced in \cite{SISOSD:JaldenOtersten:ParallelSOSD}, and further refined in \cite{SISOSD:StuderBurgBolcskei:SOSD}, for soft-output signal detection in MIMO systems, which ensures that every node in the search tree is visited at most once.
We incorporate these MIMO-SD modifications into the SD for MSDD of IR-UWB as described in \cite{UWB:Lottici:MSDD} (for a detailed description of the MIMO-SOSD, cf.\ \cite{SISOSD:StuderBurgBolcskei:SOSD}).
Thus, the required $\Lambda^{{\GLRT}}$, $\Lambda_i^{\overline{\GLRT}}$, and $\ve{a}^\GLRT$ result from a single tree search process.

First, the SD search radius $R$ is not updated, whenever a new (preliminary) best sequence has been found, but the search radius update is based on the current values $\Lambda^\GLRT$ and $\Lambda_i^{\overline{\GLRT}}$, $i=1,...,L$. It is set such, that only branches in the search tree are considered, which can lead to an update of either $\Lambda^\GLRT$, or $\Lambda_i^{\overline{\GLRT}}$, $i=1,...,L$.
This is achieved by setting
\begin{align}
        R = \max \bigg\{\max_{k = i,...,L} \Lambda_k^{\overline{\GLRT}},~\mathop{\max_{l=1,...,i-1}}_{\mathrm{with}~\tilde{a}_l \neq a_l^\GLRT} \Lambda_l^{\overline{\GLRT}}\bigg\} \;.
\label{eq:UWB:setSDsearchradius}
\end{align}
%

Further, in the case a sequence $\veat$ with path metric $\Lambda(\veat)$ is investigated, i.e., a leaf in the search tree has been reached, two cases are distinguished:\\
(i) If $\Lambda(\veat) < \Lambda^\GLRT$, a new (preliminary) best sequence has been found. Then all $\Lambda_i^{\overline{\GLRT}}$, $i=1,...,L$, where $\tilde{a}_i = -a_i^\GLRT$, are set to $\Lambda^\GLRT$, followed by the usual SD update of the current best sequence $\ve{a}^\GLRT := \veat$ and metric $\Lambda^\GLRT := \Lambda(\veat)$.\\
(ii) If $\Lambda(\veat) \geq \Lambda^\GLRT$, only the counterhypotheses have to be checked, i.e., all $\Lambda_i^{\overline{\GLRT}}$, $i=1,...,L$, where $\tilde{a}_i = -a_i^\GLRT$ and $\Lambda_i^{\overline{\GLRT}} > \Lambda(\veat)$, are set to $\Lambda(\veat)$.
%
\subsubsection{SOSD Complexity Reduction}
A reasonable measure for the search complexity of the SD is the number of visited nodes $C_{\mathsf{SD}}$ in the search tree during the tree search process, which is directly related to hardware implementation complexity, cf., e.g., \cite{SISOSD:StuderBurgBolcskei:SOSD}.
In this paper, we adopt this complexity measure.

In \cite{me:ICUWB09} we introduced a packing-radius-based stopping criterion for the SD in hard-output MSDD of IR-UWB, which reduces the average search complexity, yet ensures optimality of the estimated sequence.
We also apply this stopping criterion for the SOSD.
If any preliminary sequence $\veat$ with path metric $\Lambda(\veat)$ fulfills
\begin{align}
        \Lambda(\veat) \leq \Rstop = L \cdot \mathop{\min_{i=1,...,L}}_{l=0,...,i-1} |Z_{l,i} |
\label{eq:UWB:stopcrit}
\end{align}
the search process is terminated early.
In \cite{me:ICUWB09} it has been shown, that this setting
%
%
guarantees to find the GLRT-optimal sequence $\ve{a}^\GLRT$ and $\Lambda^\GLRT$.
However, this does not ensure to find the optimal solution to (\ref{eq:UWB:Lambdacounter}), i.e., $\Lambda_i^{\overline{\GLRT}}$, and hence the correct max-log-approximated LLRs.
However, as will be shown later, this stopping criterion enables a reduction in the average search complexity at only minor performance degradation.

As shown in \cite{SISOSD:StuderBurgBolcskei:SOSD}, a crucial part for complexity reduction of the SOSD in MIMO detection, is to limit the maximum LLR values during the SD search process.
For MIMO detection this LLR clipping enables a tradeoff between the power efficiency of optimal soft-output detection and the complexity of hard-output detection.
Since we aim for a similar tradeoff, we apply this LLR clipping with maximum LLR value $\LLRmax$ during the search process, too.
Thus, after each update of the counterhypotheses metrics $\Lambda_i^{\overline{\GLRT}}$, those are limited to
\begin{align}
        \Lambda_i^{\overline{\GLRT}} = \max\bigg\{\Lambda_i^{\overline{\GLRT}}, \Lambda^{{\GLRT}}+\sigma_n^2(L+1)\LLRmax\bigg\},\;\forall i\;.\raisetag{2mm}
\label{eq:UWB:LLRclipping}
\end{align}
From (\ref{eq:UWB:setSDsearchradius}) it can be seen that this LLR clipping limits the search radius to $R \leq \Lambda^{{\GLRT}}+\sigma_n^2(L+1)\LLRmax$ and, together with (\ref{eq:UWB:LLR_cases}), ensures that $|\LLR_i| \leq \LLRmax$ after the SD search.
Clearly, with $\LLRmax = 0$ hard-output SD-based MSDD (HOSD) results.


%
\subsubsection{SOSD Algorithm}
The resulting algorithm, based on the SD for MSDD of IR-UWB \cite{UWB:Lottici:MSDD}, is given in pseudo-code representation in Fig.~\ref{fig:UWB:pseudocode}.
The modifications for soft output generation, LLR clipping (cf.\ line {19}), and the stopping criterion (cf.\ line {15}), are highlighted in gray shading.
Note that this pseudo-code representation is based on the pseudo-code representation of the SD for MSDD of DPSK given in \cite{MSDD:Lampe:MSDSD}.
We introduced the counter $n_i$, which is used to check if the two branches emanating from each node have been checked, the branch metric
\begin{align}
        \delta = \sum_{l=0}^{i-1} \left( \left| Z_{l,i} \right| \, \left( 1 - \sign\left(Z_{l,i}\right) \, \prod_{k=l+1}^{i} a_{k}   \right) \right)
\label{eq:UWB:branchmetric}
\end{align}
and the path metric $\pathMw{i}$ at a node in depth $i$.
We used the short-hand writing $\Lambda^{\overline{\GLRT}}_i = \Lambda_i$ and $\Lmax = \sigma_n^2(L+1)\LLRmax$, and omitted the tilde for a hypothesis.
The sub-functions \texttt{findbest} and \texttt{findnext} have been introduced (f1 and f2), to account for branch enumeration according to the Schnorr-Euchner search strategy \cite{SD:Agrell:Closest}.

%
\begin{figure}[!t]
        \input{pseudocode_UWB}
\caption{Pseudo-code representation of the SOSD algorithm for soft-output MSDD of IR-UWB. Gray-shaded: modifications for soft output generation.}
\vspace*{-5mm}
\label{fig:UWB:pseudocode}
\end{figure}
%

%
\section{Performance and Complexity}
\label{sec:results}
We evaluate the performance and complexity of the proposed coded IR-UWB transmission system in a typical UWB scenario, where we assume no inter-symbol interference ($\T$ chosen sufficiently large), 
and $\pTX(t)$ is a Gaussian monocycle with $2.25\GHz$ center frequency and $10\,\dB$ bandwidth of $3.3\GHz$.
The propagation channel is modeled according to IEEE-CM\,2 \cite{UWB:Molisch:Channel} with each realization normalized to unit energy.
The receive filter is matched to the transmit pulse shape and a good compromise for the integration time is $\Ti = 30\ns$.
We use maximum-free-distance rate-1/2 convolutional codes with $2^\nu$ states, an interleaver size of 1000 information bits, and the Viterbi algorithm as (soft-input) channel decoder \cite{book:Proakis:DigitalCommunications}.

\begin{figure}[!t]
\centering
\psfrag{xlabel}[]{$10\log(\Eb/\No)~[\dB]$ $\rightarrow$}
\psfrag{ylabel}[]{$\BER$ $\rightarrow$}
\psfrag{L=2}[cl][.99]{\hspace*{-3ex}\small$L\hspace*{-.4ex}=2$}
\psfrag{L=5}[cl][.99]{\hspace*{-3ex}\small$L\hspace*{-.4ex}=5$}
\psfrag{L=10}[cl][.99]{\hspace*{-3.5ex}\small$L\hspace*{-.4ex}=\hspace*{-.3ex}10$}
\psfrag{L=15}[cl][.99]{\hspace*{-3.5ex}\small$L\hspace*{-.4ex}=\hspace*{-.3ex}15$}
\psfrag{L=20}[cl][.99]{\hspace*{-3.5ex}\small$L\hspace*{-.4ex}=\hspace*{-.3ex}20$}
\psfrag{L=25}[cl][.99]{\hspace*{-3.5ex}\small$L\hspace*{-.4ex}=\hspace*{-.3ex}25$}
        \includegraphics[width = .9\columnwidth]{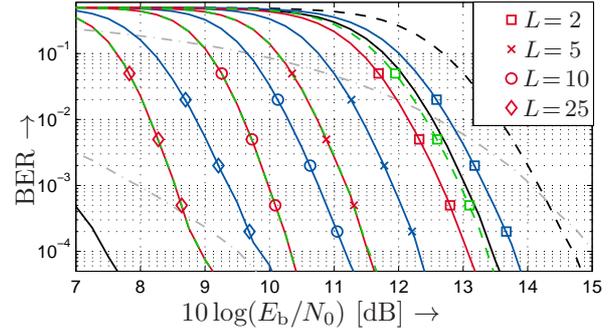}
\caption{BER vs.\ $\Eb/\No$ in $\dB$ of coded IR-UWB with soft-output MSDD using the SOSD without (red) and with stopping criterion (\ref{eq:UWB:stopcrit}) (dashed green), and hard-output MSDD (blue) for different $L$, in comparison to soft and hard DD ($L=1$, black, dashed black), coherent detection (left most, black), and uncoded transmission (coh.\ and DD: dash-dotted gray). Rate-$1/2$ conv. code with $\nu = 6$, $\LLRmax = 10$, IEEE-CM\,2.}
\vspace*{-4mm}
\label{fig:UWB:BER}
\end{figure}

Fig.~\ref{fig:UWB:BER} shows the BER performance of the coded IR-UWB system employing the proposed SOSD-based MSDD.
With increasing MSDD block size $L$ performance improves compared to DD and approaches (hard-decision) coherent detection.
Soft-output MSDD (using SOSD) achieves an additional gain of up to  $1\;\dB$ over hard-output MSDD ($\LLRmax = 0$).

\begin{figure}[!t]
\centering
\psfrag{L2}[b]{\small$L=2$}
\psfrag{L3}[b]{\small$L=3$}
\psfrag{L4}[b]{\small$L=4$}
\psfrag{L5}[b]{\small$L=5$}
\psfrag{L6}[b]{\small$L=6$}
\psfrag{L8}[bl]{\small$L=8$}
\psfrag{L10}[bl]{\small$L=10$}
\psfrag{L12}[b]{\small$L=12$}
\psfrag{L15}[b]{\small$L=15$}
\psfrag{LLRmax}[bl]{\rotatebox{-20}{\small$\leftarrow\LLRmax\hspace*{-.8ex}\uparrow$}}
\psfrag{xlabel}[]{$\Eb^{\mathsf{req}}/\No$ $[\dB]$ $\rightarrow$}
\psfrag{ylabel}[]{avg. $C_{\mathsf{SD}}$ $\rightarrow$}
        \includegraphics[width = .9\columnwidth]{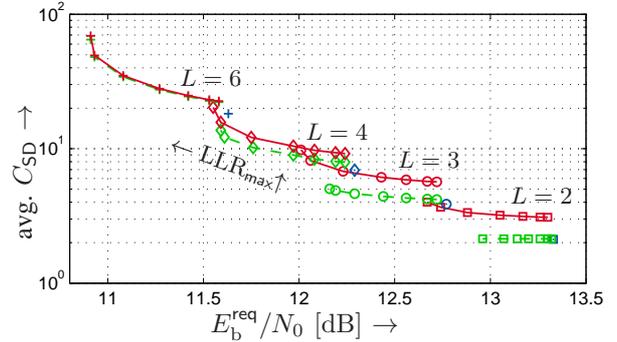}
\caption{Tradeoff performance vs.\ average SD complexity at $\BER = 10^{-3}$ for LLR clipping levels $\LLRmax = [0.05\;0.1\;0.25\;0.5\;1\;2\;10]$ (right to left) of MSDD of IR-UWB using the SOSD (red), the SOSD with stopping criterion (\ref{eq:UWB:stopcrit}) (green), and hard-output MSDD (blue) with different MSDD block sizes $L$. Rate-$1/2$ conv. code with $\nu = 6$, IEEE-CM\,2.}
\vspace*{-5mm}
\label{fig:UWB:tradeoff}
\end{figure}

Adjusting the LLR clipping level $\LLRmax$ in SOSD, a tradeoff between the power efficiency of GLRT-optimal max-log-approximated soft-output MSDD and hard-output MSDD is achieved, which almost continuously traverses the performance-complexity plane (cf.\ Fig.~\ref{fig:UWB:tradeoff}).
Additionally setting the stopping criterion (cf.\ (\ref{eq:UWB:stopcrit}) and line 15 in Fig.~\ref{fig:UWB:pseudocode}) to enable early termination of the SOSD search process, especially for small to moderate $L$ reduces the average SD complexity at only minor loss in performance.

However, using the (soft-input) Viterbi algorithm for channel decoding imposes an increase in overall receiver complexity depending mainly on the number of states $2^\nu$ of the convolutional code \cite{book:Proakis:DigitalCommunications}.
Hence, the question for the optimum
 tradeoff between power efficiency and overall receiver complexity, obtained from adjusting the major parameters code constraint length, MSDD block size, and LLR clipping level, arises.

Since the proposed SOSD requires no multiplications  (note that $\tilde{a}_i \in\{\pm1\}$), we assume that soft-input Viterbi decoding and the SOSD for MSDD are comparable with respect to complexity in the number of considered nodes in the trellis, respectively the binary search tree.
Hence, the overall complexity per processed symbol is $C_{\mathsf{o}} = 2^{\nu} + C_{\mathsf{SD}}/L$, where the first summand is the (fixed) complexity of the Viterbi algorithm and the second summand represents the (varying) SD complexity.
As the SD complexity depends on the ACR output, we denote the average and maximum overall complexity as $C_{\mathsf{o}}^{\mathsf{soft}}$ and $C_{\mathsf{o}}^{\mathsf{max}}$, respectively.
The worst-case or maximum SD complexity is $C_{\mathsf{SD}}^{\mathsf{max}} = \sum_{i=1}^{L} 2^i = 2^{L+1}-2$.

We compare the proposed soft-output MSDD system with a reference system employing a $2^{\nu_{\mathsf{ref}}}$-states convolutional code and symbol-by-symbol ACR-based DD (MSDD with $L=1$) with overall complexity $C_{\mathsf{o}}^{\mathsf{ref}} = 2^{\nu_{\mathsf{ref}}} + 1$.
For soft-output MSDD, at each block size the best setting $\nu,\LLRmax$ is chosen (i.e., lowest $\EbNo$ for a desired $\BER$), that still has less overall complexity than the reference system (i.e., $C_{\mathsf{o}}^{\mathsf{soft}} \leq C_{\mathsf{o}}^{\mathsf{ref}}$).
For $\nu_{\mathsf{ref}} = 7$ and $\BER = 10^{-3}$ this results in the trajectories 
shown in Fig.~\ref{fig:UWB:ctotal} (chosen setting indicated 
in top part, similar results are obtained for other references $C_{\mathsf{o}}^{\mathsf{ref}}$).
The bottom part depicts the average (solid) and maximum (dashed) overall complexity using the SOSD (red), and the reference $C_{\mathsf{o}}^{\mathsf{ref}}$ (dashed black).
The top part shows the corresponding required $\Eb/\No$ in $\dB$. For comparison hard-output MSDD is included (blue).

The proposed soft-output MSDD in combination with convolutional codes with $2^\nu < 2^{\nu_{\mathsf{ref}}}$ states can---up to MSDD block sizes $L\leq16$---beat the reference system, i.e., DD and $\nu_{\mathsf{ref}} =7$.
The lowest $\Eb/\No$, with approximately $4\;\dB$ gain over the reference, is achieved for $L=12$.
However, for $L\geq15$ the complexity of the reference system is only undercut by falling back to hard-decision decoding ($\LLRmax = 0$).
With respect to the maximum overall complexity, soft-output MSDD is only better than the reference system up to $L=8$.

As for each MSDD block size the required $\Eb/\No$ of hard-output MSDD is higher than that of soft-output MSDD, at lower average, but equal maximum complexity, hard-output MSDD does not utilize the (anyway to be reserved) maximum overall complexity as good as soft-output MSDD.

\newcommand{\labelnulmax}[2]%
{${#1}\hspace*{-.4ex},\hspace*{-.4ex}{#2}\hspace*{-.0ex}$}
\begin{figure}[!t]
\centering
\psfrag{xlabel}[]{MSDD block size $L$ $\rightarrow$}
\psfrag{y1}[]{$\Eb^{\mathsf{req}}/\No$ $[\dB]$}
\psfrag{y2}[]{$C_{\mathsf{o}}$}
\psfrag{hosdc}[bl]{\small\sffamily $C_{\mathsf{o}}^{\mathsf{hard}}$}
\psfrag{sosdc}[bl]{\small\sffamily $C_{\mathsf{o}}^{\mathsf{soft}}$}
\psfrag{hosd}[bl]{\small$\mathsf{hard}$
\scriptsize($2\hspace*{-.4ex}\leq\hspace*{-.4ex}L\hspace*{-.4ex}\leq\hspace*{-.4ex}14$: \labelnulmax{6}{0})}
\psfrag{sosd}[bl]{\small$\mathsf{soft}$}
\psfrag{hodd}[cl]{\scriptsize\raisebox{.6mm}{{\tiny$<$}\hspace*{-1ex}\rule[1.1pt]{7mm}{.2pt}}\labelnulmax{7}{0}}
\psfrag{sodd}[bl]{\scriptsize\raisebox{0mm}{{\tiny$<$}\hspace*{-1ex}\rule[1.1pt]{11mm}{.2pt}}\labelnulmax{7}{\infty}}
\psfrag{6,10}[]{\raisebox{1mm}{\scriptsize\hspace*{1ex}\labelnulmax{6}{10}}}
\psfrag{l}[]{\rotatebox{-14}{\scriptsize$\rightarrow$}}
\psfrag{r}[]{\rotatebox{-5}{\scriptsize$\leftarrow$}}
\psfrag{6,1}[l]{\scriptsize\raisebox{2mm}{\hspace*{1mm}\labelnulmax{6}{1}}}
\psfrag{6,2}[l]{\scriptsize\raisebox{2mm}{\labelnulmax{6}{2}}}
\psfrag{3,0.1}[l]{\scriptsize\labelnulmax{3}{\hspace*{-.4ex}\frac{1}{10}}}
\psfrag{6,0}[l]{\scriptsize\raisebox{2mm}{\labelnulmax{6}{0}}}
\psfrag{5,2}[l]{\scriptsize\raisebox{2mm}{\labelnulmax{5}{2}}}
\psfrag{5,1}[l]{\scriptsize\raisebox{2mm}{\labelnulmax{5}{1}}}
\psfrag{4,0.5}[l]{\scriptsize\raisebox{2mm}{\labelnulmax{4}{\hspace*{-.4ex}\frac{1}{2}}}}
\psfrag{6,0.25}[l]{\scriptsize\raisebox{2mm}{\labelnulmax{6}{\hspace*{-.4ex}\frac{1}{4}}}}
\psfrag{5,0.25}[l]{\scriptsize\raisebox{2mm}{\labelnulmax{5}{\hspace*{-.4ex}\frac{1}{4}}}}
\psfrag{4,0}[l]{\scriptsize\raisebox{2mm}{\labelnulmax{4}{0}}}
\psfrag{5,0}[l]{\scriptsize\raisebox{2mm}{\labelnulmax{5}{0}}}
\psfrag{(6)}[]{\scriptsize}
\psfrag{(5)}[l]{}
\psfrag{(4)}[l]{}
\psfrag{dd}[bl]{\small\sffamily $C_{\mathsf{o}}^{\mathsf{ref}}$}
\psfrag{cmax}[bl]{\small\sffamily $C_{\mathsf{o}}^{\mathsf{max}}$}
        \includegraphics[width = .99\columnwidth]{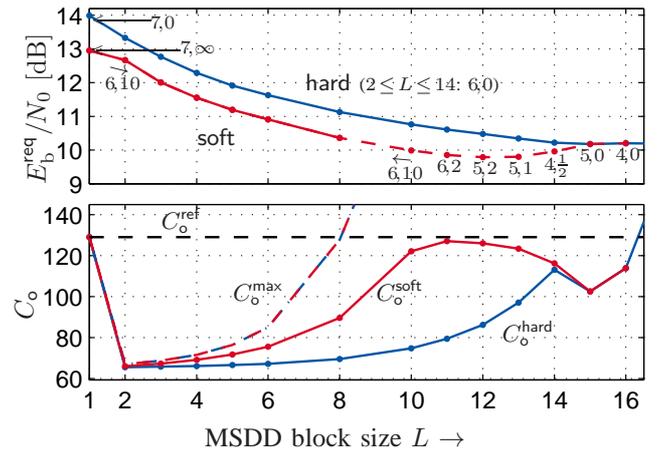}
\caption{Trajectories of performance and overall receiver complexity of the soft-output and hard-output MSDD IR-UWB receiver in comparison to a DD reference with $\nu_{\mathsf{ref}} = 7$ ($C_{\mathsf{o}}^{\mathsf{ref}} = 2^7+1$) at $\BER = 10^{-3}$. Lables (\labelnulmax{\nu}{\hspace*{.4ex}\LLRmax}) indicate chosen setting for soft-output MSDD (see text). Rate-1/2 conv.\ codes with $2^\nu$ states, IEEE-CM\,2.}
\vspace*{-5mm}
\label{fig:UWB:ctotal}
\end{figure}
%
%
\section{Conclusions}
\label{sec:conclusions}
We have presented a noncoherent SD-based soft-output MSDD receiver for coded IR-UWB transmission systems.
Based on the GLRT approach, we have derived the LLRs and formulated their computation as tree search problems, enabling the application of the SD for efficient implementation.
Incorporating recent results from MIMO detection, we are able to compute the LLRs in a single SD tree search.
In combination with further techniques for SD complexity reduction, the proposed soft-output SD for MSDD of IR-UWB thus imposes only a moderate complexity increase compared to hard-output MSDD.
Employing this soft-output demodulator for IR-UWB, a large part of the gap between conventional noncoherent DD and ideal coherent detection can be closed.

%
%
%
%


%
%
%
%

%
\end{document}